\documentclass[twoside,twocolumn,english,pra,showpacs]{revtex4}
\usepackage{babel}
\usepackage{graphicx}
\usepackage{dcolumn}
\usepackage{amsmath}
\usepackage{float}

\begin{document}
       
\title{Photon-pair production at the nanoscale with hybrid nonlinear/plasmonic antennas}

\author{G. Laurent, N. Chauvet, G. Nogues, A. Drezet and G. Bachelier}
\affiliation{Univ. Grenoble Alpes, CNRS, Institut N\'eel,
38000 Grenoble, France}

\begin{abstract}
Integration of photon pairs nanosources is a major challenge for quantum technology. In this context, we develop a formalism for the investigation of Spontaneous Parametric Down Conversion in hybrid structures that combine plasmonic resonances and intrinsic nonlinearity. Using quantum and numerical approaches together, we quantitatively evaluate photon pair correlation measurements for realistic experimental configurations. Hybrid structures embedding a small nanocystal allow for a $10^{3}$ fold increase in pair production compared to the same nanocrystal alone and a photon pair production efficiency close to the best source to date.
\end{abstract}


\maketitle

\section{Introduction}

Manipulating entangled states has appeared to be a promising way for entering into the age of quantum cryptography \cite{Ekert1991}, fostering the development of bright and reliable photonic sources, initially based on spontaneous parametric down conversion (SPDC) \cite{Kwiat1995} and producing nowadays almost pure entangled states with high efficiency \cite{Weston2016}. Meanwhile, a variety of strategies has been proposed for producing/managing/analyzing entangled photon states \cite{Pan2012}, even over global distances via satellite links \cite{Boone2015} or with many particles \cite{Wang2016}. But in view of integration, the major challenge is to provide micro/nano-scale building block devices pertaining high figures of merit such as high degree of entanglement, large brightness and some degree of tunablility \cite{Orieux2017}. With these respects, quantum dots (QD) appear as promising candidates \cite{Akopian2006,Jayakumar2013,Kuroda2013}, especially when they are embedded in photonic structures \cite{Dousse2010,Versteegh2014,Jons2017} for enhanced Purcell factors, in piezoelectric host \cite{Trotta2014} to reduce the natural strain-induced finite structure splitting of the excited states or in gated environment \cite{Zhang2017,CHen2016} for Stark-effect-driven tuning over a few meV. The utmost advantage of QD over SPDC-based sources resides in their ability to provide triggered entangled photons \cite{Hafenbrak2017,Zhang2015} and even on-demand and coherent emission using resonant two-photon excitation \cite{Muller2014,Huber2016}. Nevertheless, QD have not yet reached the level of entanglement purity achieved with macroscopic SPDC sources and suffer from a major drawback: the need for cryogenic environment \cite{Orieux2017}. 

On the other hand, SPDC based sources have been extensively used thanks to their uncompeted degree of entanglement allowing to test the principles of quantum mechanics by demonstrating the Hong-Ou-Mandel photon interference effect \cite{Hong1987}, the entanglement swapping \cite{Pan1998} or more recently loophole-free tests of Bell's Theorem \cite{Giustina2015, Shalm2015}. However, to our knowledge, no attempt has been made for investigating SPDC in nanosized crystals. The main reason is the dramatic decrease of efficiency while reducing the nonlinear medium size. Clearly, the latter requires compensation by enhancing the pump excitation or the photon emission efficiency. This can be achived by boosting the local electromagnetic fields with optical cavities \cite{Oberparleiter2000} or plasmonic antennas \cite{Nevet2010,Poddubny2012,Poddubny2016}. Although this second option allows building compact (truely nanosized) structures, capable of polarization control \cite{Maksymov2012}, it requires novel tools for evaluating and optimizing the overall structure efficiency.

Various approaches have been followed to simulate photon pair generation, including (i) density matrix formulations for resonant electronic excitations, such as in biexciton-exciton cascades \cite{Larque2008}, in superconducting circuits \cite{Moon2005,Marquardt2007} or in qubit-plasmonic antenna coupled systems \cite{Hou2014,Straubel2017} and (ii) hamiltonian treatments, provided that the mathematical expression of the optical modes is known \cite{Hong1985,Ling2008, Vernon2015}. The latter approach is, however, no longer valid for nanostructures holding a nonlinear crystal due to their complex shapes. To circumvent this issue and account for the vacuum field fluctuations in quantum treatments, one can use a stochastic model introducing a white input noise triggering the process \cite{Brambilla2004}, or express only the measured correlations, from which the quantum fluctuations disappear in favor of classical quantities \cite{Mitchell2009}.

In this work, we present an approach mixing quantum formalism and classical numerical simulations for modeling SPDC in nanosized systems. We develop a theoretical expression for photon pair correlations, under periodically pulsed excitation, that does not depend on a quantum vacuum fluctuation term but only on classical green functions, correlating source and detector electric fields. Thanks to the numerical modeling of the near-field electromagnetic response, the photon pair production rate is quantitatively evaluated in nanostructures with no analytical mode, such as a nonlinear crystal coupled to plasmonic antennas. We investigate the feasibility and the optimization of photon-pair production, opening avenues for entanglement management in SPDC based sources at the nanoscale.

\section{Model}
In the framework of quantum measure theory, the probability for simultaneously detecting two photon with linear polarizations $\alpha_{1/2}$ at location $\mathbf{r}_{1/2}$ and time $t_{1/2}$ is \cite{Glauber1963}
\begin{equation}\label{probaDetect}
\begin{split}
G(\alpha_{1},\mathbf{r_{1}},&t_{1},\alpha_{2},\mathbf{r_{2}},t_{2}) \propto\\ &\left|\left<\Psi_f\right|\,\hat{E}^{(+)}_{\alpha_{1}}(\mathbf{r_{1}},t_{1})\hat{E}^{(+)}_{\alpha_{2}}(\mathbf{r_{2}},t_{2})\,\left|\Psi_{int}\right>\right|^2.
\end{split}
\end{equation}
This quantity depends on the photon state before $\left|\Psi_\text{int} \right>$ and after $\left|\Psi_\text{f} \right>$ detection, and on the electric field operators $\hat{E}^{(+)}_{\alpha}(\mathbf{r},t)$. Assuming a weakly depleted pump regime, we adopt a perturbative approach in the interaction representation at the first order, so that $\left|\Psi_{int}\right>$ is given by: 
\begin{equation}\label{perturbation}
\left|\Psi_{int}\right> \simeq \left(1 - \frac{i}{\hbar} \int_{0}^{T_r}\mathcal{\hat{H}}_{int} dt\right)\left|\Psi_{i}\right>
\end{equation}
with
\begin{equation}\label{hamiltonian}
\begin{split}
&\mathcal{\hat{H}}_{int} = -\int_{V} \mathbf{\hat{E}}^{(+)}_p(\mathbf{r}_{n},t)\cdot \\ 
&\varepsilon_{0}\overset{\leftrightarrow}{\chi}^{(2)}\colon \mathbf{\hat{E}}^{(-)}_s(\mathbf{r}_{n},t) \mathbf{\hat{E}}^{(-)}_i(\mathbf{r}_{n},t)\ d^3r_{n} + \text{h.c.c.}.
\end{split}
\end{equation}
$\left|\Psi_{i}\right>$ is the initial state and $\mathcal{\hat{H}}_{int}$ corresponds to the dipolar type interaction Hamiltonian written in the rotating wave approximation and involving the second order nonlinear tensor $\overset{\leftrightarrow}{\chi}^{(2)}$ of the nanocrystal. Although this calculation describes the optical response of a lossy medium through a Hamiltonian approach, it has been shown to be robust \cite{Scheel2006}. 

\begin{figure} \center 
{\includegraphics[width = 0.5\textwidth]{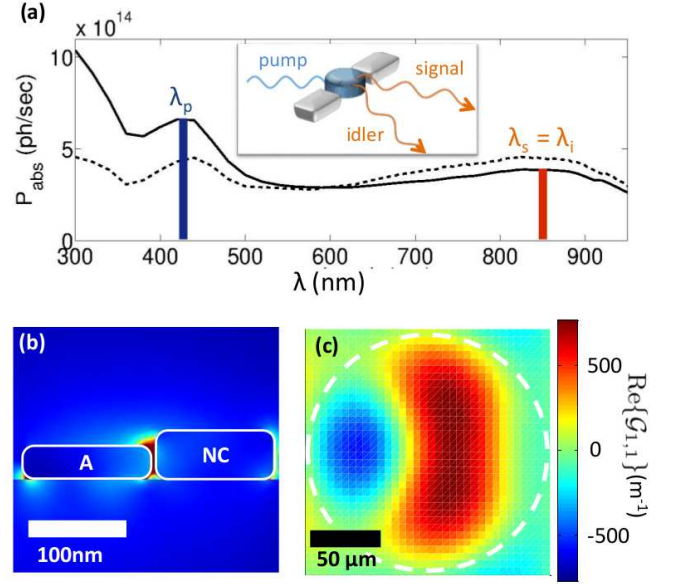}}
\caption{(a) Simulated absorption spectra of two antennas (dashed line) and a hybrid structure (plain line) holding a KTP crystal (see inset)resulting from a dipolar excitation at the center of the two antenna. (b) Pump electric field norm at $\lambda = 425$ nm for a hybrid structure composed of a 130$\times$100$\times$35nm$^3$ aluminium antenna (A) and a \mbox{130 nm} wide--\mbox{50 nm} high KTP cylindrical crystal (NC). (c) $xx$ real component of the Green's function, labeled $\text{Re}\{\mathcal{G}_{1,1}\}$ and associated with a dipole located at the center of the crystal ($\lambda=$ \mbox{850 nm}) for different detection points. The dashed circle corresponds to the edge of a \mbox{170 $\mu$m} wide circular detector.\label{pump}}
\end{figure}

The system is assumed to be described by a Fock state $\left|N_p,N_s,N_i\right>$ where $N_p$ (resp. $N_s$ or $N_i$) is the photon number at pulsation $\omega_p$ (resp.  $\omega_s$ or $\omega_i$), $p$ referring to the \emph{pump}, $s$ and $i$ to the \emph{signal} and \textit{idler} down-converted photons. Those numbers are linked by the Feynman diagram illustrated in inset of Fig.~\ref{pump}(a): after interaction, the system is in a superposition of the two states $\left|N_p,0,0\right>$ and $\left|N_p-1,1,1\right>$. 

The number $\mathcal{N}_\text{corr}$ of correlated photons measured on the two detectors can be quantitatively evaluated from Eqs.~(\ref{probaDetect}-\ref{hamiltonian}). The calculation follows several stages detailed in the Supplemental Material and summarized as follows: 

\textit{(i)} We assume a pulsed excitation with a repetition rate $\omega_r$, so that the pump electric field and the SPDC response fields can be developed in Fourier series
\begin{equation}
\mathbf{\hat{E}}(\mathbf{r},t) = \sum_{\omega=\omega_r}^{\infty} \mathbf{\hat{E}}(\mathbf{r},\omega) e^{-i\omega t}
\end{equation}
with $\omega$ a multiple of $\omega_r$. 

\textit{(ii)} The excitation pump is assumed to be in a coherent state. Hence, the corresponding Fourier coefficients are treated as components of a classical field $\mathbf{E}_p$ instead of quantum operators $\mathbf{\hat{E}}_p$. Moreover, as shown in Fig.~\ref{pump}(b) for a hybrid structure holding a KTiOPO$_{4}$ (KTP) crystal located near an aluminum antenna, the electric field is strongly localized in the vicinity of the metal surfaces. Numerical simulations of the scattered near-field at the pump wavelength are thus necessary to properly account for the optical response of the structure \cite{Ethis-de-Corny2016}.

\textit{(iii)} Creation and destruction operators for signal and idler appear in paired quantities such as $\left<0\right|\hat{E}^{(+)}_{\alpha_k}(\mathbf{r}^{\prime},\omega)\hat{E}^{(-)}_{\alpha_j}(\mathbf{r},\omega)\left|0\right>$ corresponding to the correlation of the quantum noise between the two locations $\mathbf{r}$ and $\mathbf{r}^{\prime}$, as no initial photon is present at these frequencies. Although these correlators derive from a quantum formalism, they are related to Green functions $\mathcal{G}^{\omega}_{\alpha_k,\alpha_j}(\mathbf{r}^{\prime},\mathbf{r})$ \cite{Mitchell2009}, that are classical entities linking the electric field at location $\mathbf{r}^{\prime}$ to an oscillating dipole at location $\mathbf{r}$:
\begin{equation}\label{correlator}
\left<0\right|\hat{E}^{(+)}_{\alpha_k}(\mathbf{r}^{\prime},\omega)\hat{E}^{(-)}_{\alpha_j}(\mathbf{r},\omega)\left|0\right> = \frac{\hbar \omega^{2}}{i\epsilon_{0}c^{2}}\mathcal{G}^{\omega}_{\alpha_k,\alpha_j}(\mathbf{r}^{\prime},\mathbf{r})\frac{1}{T_r}.
\end{equation}
The detailed calculation \cite{Mitchell2009} is adapted here for Fourier series. Fig.~\ref{pump}(c) shows an example of $xx$ real component of the Green's function, labeled $\text{Re}\{\mathcal{G}_{1,1}\}$, for a dipole radiating at the center of the KTP crystal (see Fig.~\ref{pump}(b)). Its strong spatial variation over the detector surface emphasizes the need for a full description of the collection path.

\textit{(iv)} The number of correlation is time-integrated over the acquisition time $T_\text{ac}$, and space integrated over the detector surfaces $S_{1/2}$. Furthermore, we consider that there is no correlation between successive pulses (\textit{i.e.} no memory effect).

\textit{(v)} The excitation spectrum is assumed to be a rectangular function of width $\Delta \omega_p$ centered on $\overline{\omega}_p$ in order to account for the limited pulse duration. We also introduce spectral selections at wavelengths $\overline{\omega}_1$ and $\overline{\omega}_2 = \overline{\omega}_p - \overline{\omega}_1$ with $\Delta \omega_1$ and $\Delta\omega_2$ widths.

\noindent The final expression for the number of detected photon-pairs is given by
\begin{widetext}
\begin{equation}\label{Correlation3}
\begin{split}
\mathcal{N}_\text{corr}(\alpha_{1},\alpha_{2}) = & T_{ac} \Delta f_{1} \frac{ \overline{\omega}_{1}^{3}(\overline{\omega}_p-\overline{\omega}_1)^{3}}{4c^{6}} \left|\tau_{\alpha_1}(\overline{\omega}_{1}) \tau_{\alpha_2}(\overline{\omega}_p-\overline{\omega}_{1})\right|^2 \\
&\times \int_{S_1}\int_{S_2} \left| \sum_{\alpha_p, \alpha_s, \alpha_i} A_{\alpha_p, \alpha_s, \alpha_i}^{\overline{\omega}_p,\overline{\omega}_1}(\mathbf{r}_1 , \mathbf{r}_2) \right|^2 d^2r_{1} d^2r_{2}
\end{split}
\end{equation}
with
\begin{equation}\label{amplitude}
A_{\alpha_p, \alpha_s, \alpha_i}^{\overline{\omega}_p,\overline{\omega}_1}(\mathbf{r}_1 , \mathbf{r}_2) = \int_{V} E^{\text{cw}}_{\alpha_p}(\mathbf{r_{n}},\overline{\omega}_p)\chi^{(2)}_{\alpha_p, \alpha_s, \alpha_i}\mathcal{G}^{\overline{\omega}_{1}}_{\alpha_{1},\alpha_s}(\mathbf{r}_{1},\mathbf{r}_{n})
\mathcal{G}^{\overline{\omega}_p-\overline{\omega}_{1}}_{\alpha_{2},\alpha_i}(\mathbf{r}_{2},\mathbf{r}_{n})\ d^3r_{n}.
\end{equation}
\end{widetext}
where we can separate two important features: \textit{the ability to generate photon pairs}, related to the pump field and $\chi^{(2)}$ amplitude, and \textit{the ability to radiate the down-converted photons to the far-field}, quantified by the Green functions, both being classical quantities, which can be numerically evaluated.

\section{Results and discussion}

In the following, we investigate the joint enhancement of these phenomena for a specific but realistic configuration: a KTP nonlinear crystal surrounded by one or two plasmonic antennas (referenced as the \textit{hybrid structure} in this work). The electric field in the gap between the antennas being strongly oriented along the antenna axis, for longitudinal orientation \cite{Novotny2012}, we restrain the study to the case where the main nonlinearity of the KTP crystal is oriented along the antennas axis. The excitation consists in a pulsed laser (average power of 100~$\mu$W, repetition rate $f_{r} = $~80~MHz, pulse duration $\Delta t_p =$~ 100~fs, $\lambda_p = $ 425 nm) focused on the sample through a high numerical aperture (NA = 1.3) immersion oil objective. The gass substrate has a matched refractive index $n = 1.518$. The shape of each antenna is tuned by varying its length in order to obtain plasmonic resonances at both $\lambda_p = 425 \text{ nm}$ and $\lambda_s = \lambda_i = 850 \text{ nm}$ (see Fig.~\ref{pump}(a)). The metal is chosen to be aluminum because of the presence of plasmonic resonances in the entire visible range, yielding doubly resonant conditions \cite{Ethis-de-Corny2016,Celebrano2015}. In order to reproduce realistic experimental configurations, the propagation of the excitation beam through the objective and the substrate is analytically implemented \cite{Novotny2012}. The currents generated by the pump field are computed using a finite elements method allowing to retrieve the near-field distribution in an auto-consistent way \cite{Ethis-de-Corny2016}. 

To model the SPDC source term, we consider dipoles located inside the nonlinear crystal. For each location $\mathbf{r}_n$ and each orientation $\beta$, we numerically compute the radiated electric field $E^\text{cw}_{\alpha}(\mathbf{r},\omega)$ on a 170~$\mu$m wide detector and evaluate the green function $\mathcal{G}^{\omega}_{\alpha,\beta}(\mathbf{r},\mathbf{r}_{n}) = i E^\text{cw}_{\alpha}(\mathbf{r},\omega) \lambda c \varepsilon_{0}/2 \pi$ \cite{Novotny2012}. The propagation of the down converted electromagnetic field through the collection objective and a focusing lens is analytically realized as for the focusing (see Ref.~\cite{Novotny2012}).

\begin{figure} \center 
{\includegraphics[width = 0.5\textwidth]{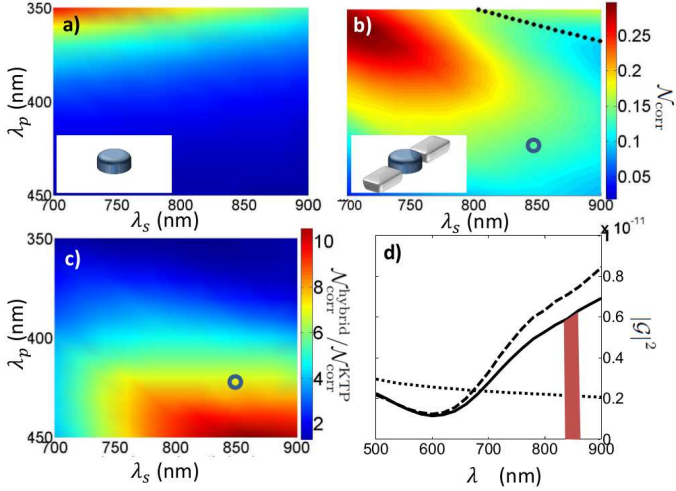}}
\caption{Number of correlations generated by two different structures: (a) an isolated KTP crystal and (b) a KTP crystal in the gap between two aluminum antennas. The open circle corresponds to the doubly resonant regime \mbox{$\lambda_p$ = 425~nm}, \mbox{$\lambda_s$~= $\lambda_i$~=~850~nm}. The dotted line corresponds to the number of correlations with a given idler wavelength $\lambda_i$ = 620~nm. (c) Enhancement factor corresponding to $\mathcal{N}^\text{hybrid}_\text{corr}/\mathcal{N}^\text{KTP}_\text{corr}$. (d) Green's function squared amplitude integrated over the detector surface ($\left|\mathcal{G}\right|^2$, see text) for three configurations: the isolated KTP crystal (dotted line), the antennas (dashed line) and the hybrid structure (full line). The wavelength labeled in red corresponds to the signal and idler wavelengths (850~nm).\label{Ncorr}}
\end{figure}

For sake of simplicity, we consider in a first step only one dipole centered on the nonlinear crystal and evaluate the dependency of the correlation number with respect to the pump and signal wavelengths (spectral integration over $\Delta\lambda_1 =$ 10~nm) for two configurations: an isolated 120 nm wide--50 nm high cylindrical KTP crystal and the same KTP crystal coupled to two 130~nm~$\times$ 100~nm~$\times$~35~nm aluminum antennas, see Figs.~\ref{Ncorr}(a-b). The number of correlations for the bare crystal smoothly decreases with the pump and signal wavelengths, while it exhibits a more complex behavior for the hybrid structure. In contrast with bulky materials \cite{Hong1985} there is no phase matching here due to the nanosized crystal. However, the addition of aluminum antennas sustaining plasmonic resonances strongly tailors the spectral response so that the standard phase matching condition is replaced by a resonant mode matching \cite{Ethis-de-Corny2016,Celebrano2015}. This is evidenced in Fig.~\ref{Ncorr}(c): the ratio between the correlation numbers measured with the hybrid structure and with the isolated KTP shows that the SPDC response is increased by one order of magnitude for pump and signal wavelengths close to resonance, further emphasizing the ability to engineer the desired spectral enhancement.

Part of the SPDC enhancement originates from the capacity of the hybrid structure to radiate photons. As we consider a KTP crystal oriented along the antenna axis, the main dipolar contribution to the down-converted fields is also oriented in this direction. As a consequence, the radiated fields mainly depend on three components of the Green function: $\mathcal{G}_{1,1}$, $\mathcal{G}_{2,1}$ and $\mathcal{G}_{3,1}$, the latter being generally negligible in far-field. The contribution of the Green functions to the overall enhancement can therefore be weighted by $\int_{S_1} \sum_{\alpha_1} \left|\mathcal{G}_{\alpha_1,1}(r_1,r_n)\right|^2 d^2r_1$ (written $\left|\mathcal{G}\right|^2$ hereafter for sake of simplicity). It is shown in Fig.~\ref{Ncorr}(d) for an isolated KTP (dotted line) and a hybrid structure (full line). The slow decay for an isolated KTP explains the results shown in the Fig.~\ref{Ncorr}(a). When considering the hybrid structure, it is worth noticing that the $\left|\mathcal{G}\right|^2$ amplitude is larger than that of a bare KTP crystal for wavelengths larger than 700~nm. Namely, the plasmonic structures fully play their role of nanoantennas, efficiently coupling the nanoscale to the far-field. Considering the degenerate case where the signal and the idler are set at 850~nm, the SPDC efficiency of the hybrid structure is 6-fold larger than for the isolated KTP crystal, see Fig.~\ref{Ncorr}(c). On the opposite, when the idler wavelength is fixed at 620~nm (deep in the dotted line of Fig.~\ref{Ncorr}(b)) the reduced Green function induces a SPDC rate drop, emphasizing here again how the SPDC efficiency can be tailored by properly designing the plasmonic resonances.

\begin{figure} \center 
{\includegraphics[width = 0.5\textwidth]{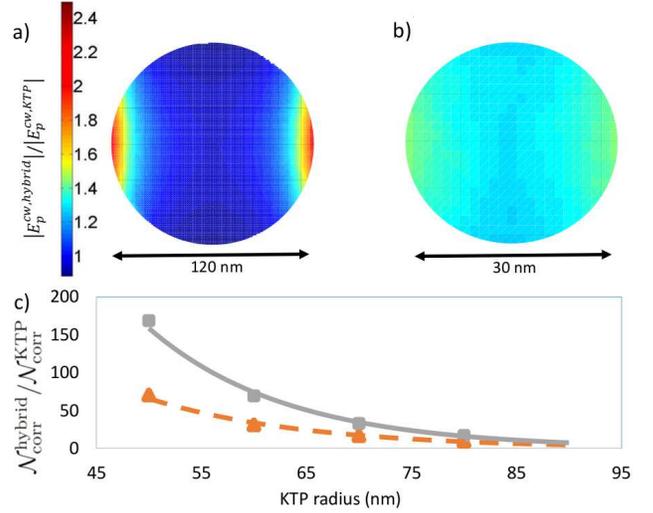}}
\caption{Absolute value of the near-field amplitude enhancement in (a) a 120~nm wide KTP and (b) a 30~nm wide KTP crystal at $\lambda_p = 425$ nm. The computation plan is parallel to the substrate and the antennas are located at the right and at the left of the crystal, with a 5~nm gap. (c) SPDC enhancement for simply-resonant gold (squares) and doubly-resonant aluminum (triangles) hybrid structures. The lines are guides to the eyes. \label{nearField}}
\end{figure}

The role of the pump field enhancement at \mbox{$\lambda_p = 425~\text{ nm}$} to the down conversion process can be evaluated by comparing the net SPDC enhancement $\mathcal{N}_\text{corr}^\text{hybrid}/\mathcal{N}_\text{corr}^\text{KTP} \simeq 6.7$ to the ratio $|\mathcal{G}_s^\text{hybrid}|^2|\mathcal{G}_i^\text{hybrid}|^2/|\mathcal{G}_s^\text{KTP}|^2|\mathcal{G}_i^\text{KTP}|^2 \simeq 7.0$. This result suggests that the pump field enhancement has a moderate (or even negative) effect on the correlation number, at least if only one effective source dipole is placed at the center of the crystal. To check the validity of this crude simplification, we mapped in Figs.~\ref{nearField}(a-b) the ratio between the $x$ component of the pump near-field in the hybrid structure and in the isolated KTP crystal for a 120~nm and a 30~nm wide--50~nm high KTP crystal. Although the field enhancement in the former is about 2.5 near the antennas (left and right sides), it is close to or even below unity at the center of the crystal. This motivated us to consider a population of $\sim$ 500 dipoles homogeneously distributed in the volume of the 120~nm wide crystal. The corresponding correlation number happens, indeed, to be reduced when compared to a single dipole source. In simple words, most of the KTP materials is too far from the antennas to benefit from the plasmonic response. Clearly, a better overlap is expected for smaller crystals, as shown in Fig.~\ref{nearField}(c) where the SPDC enhancement factor is evaluated for different KTP crystal sizes in hybrid structures made of simply-resonant gold and doubly-resonant aluminum antenna (resp. 110 nm and 130 nm long with constant 5 nm gaps between the KTP crystal and the antennas). This is consistent with the exponentially decaying near-fields at the vicinity of the metal surfaces. It results in a SPDC enhancement up to 170 for 100 nm wide KTP crystal and over 1500 for 30 nm wide KTP crystal (not represented here). This relative SPDC enhancement is several orders of magnitude higher when compared with microscale nonlinear crystal coupled to optical cavities \cite{Oberparleiter2000, Yang2007}. Yet, even for a 30 nm wide KTP crystal, the computed SPDC enhancement with a single dipole is $\mathcal{N}_\text{corr}^\text{hybrid}/\mathcal{N}_\text{corr}^\text{KTP} \simeq 1.5\times 10^{3}$ and the ratio $|\mathcal{G}_s^\text{hybrid}|^2|\mathcal{G}_i^\text{hybrid}|^2/|\mathcal{G}_s^\text{KTP}|^2|\mathcal{G}_i^\text{KTP}|^2 \simeq 1.0\times 10^{3}$. It corresponds to a pump enhancement of a few tens of percent as shown in Fig.~\ref{nearField}b, so that it is still not optimal (other materials like silver may be of interest as they show strong plasmonic resonances at the excitation wavelength and no interband transitions). Nevertheless, we would like to comment here the absolute magnitude of the SPDC rate shown in Figs.~\ref{Ncorr}(a-b). Although less than one correlation per second may sound weak, one has to fairly compare the source efficiencies by normalizing the pair generation rate by the excitation power and the squared volume of the overall device. Doing so, one gets roughly $6\times10^6$ pairs/s/mm$^6$/mW for a macroscopic Sagnac-loop based source \cite{Weston2016} and up to $2\times10^{24}$ pairs/s/mm$^6$/mW for a single QD in a Bragg cavity \cite{Dousse2010}. With this in mind, the present hybrid nanostructure yields $5\times10^{23}$ pairs/s/mm$^6$/mW, i.e. a quantum efficiency close to the best source achieved \cite{Orieux2017}.

\section{Conclusion}

In conclusion, we have developed a theoretical approach for spontaneous parametric down conversion at the nanoscale. We quantitatively evaluate the number of correlated photon pairs generated by structures composed of a nonlinear nanocrystal coupled to plasmonic antennas. We emphasize the key role of plasmonic resonances on near-field enhancement and far-field coupling. They tailor the spectral response of the hybrid nanostructures, where phase-matching conditions no longer hold. Finally, we demonstrate that photon pair generation is achievable at the single particle level, opening routes for designing compact two photon sources. We predict that resonant excitation conditions induce an enhancement of the number of correlations up to three order of magnitude when compared to SPDC from isolated KTP nanocrystal and a photon pair production efficiency comparable to that of the brightest source reported so far.

\begin{acknowledgments}
The authors acknowledge the financial support from the Agence Nationale de la Recherche, France, through the TWIN project (Grant No. ANR-14-CE26-0001-01-TWIN), the Universit\'e Grenoble Alpes, France, for the Chaire IUA award to G.B and the Ph.D. grant to N.C. from the Laboratoire d?excellence LANEF in Grenoble (ANR-10-LABX-51-01).
\end{acknowledgments}

\end{document}